**Kinesthetic activities in physics instruction: Image schematic justification and design based on didactic situations**

Jesper Bruun and Frederik V. Christiansen

*Abstract*
One of the major difficulties in learning physics is for students to develop a conceptual understanding of the core concepts of physics. Many authors have argued that students' conceptions of basic physical phenomena are rooted in basic schemas, originating in fundamental kinesthetic experiences of being. If central schemas have a bodily basis, this idea should be utilized in physics instruction. Thus, we argue that kinesthetic activities, including careful experiential and conceptual analysis will provide useful entry point for students' acquisition of the basic conceptions of physics, and can overcome the phenomenological gap between the experiential and the conceptual understanding. We discuss the nature of image schemas and focus particularly on one: *effort-resistance-flow*. We argue that this schema is fundamental not only in our everyday experience, but also in most of school physics. We provide an example of a kinesthetic model and describe how an instructional strategy of these exercises can support student understanding and intuition with respect to central physics concepts.

En af vanskelighederne ved at lære fysik er at de studerende skal tilegne sig en sammenhængende begrebslig forståelse af fysikkens grundbegreber. Mange forfattere har understreget, at studerendes begrebslige forståelse af grundlæggende fysiske fænomener har rod i mentale skemaer, der har udspring i kropslige erfaringer. Givet at dette er tilfældet, bør det udnyttes i fysikundervisningen. Vi diskuterer hvordan de studerendes skemaer kan formodes at være organiseret, og vil fokusere på et særligt vigtigt skema, det såkaldt "anstrengelse-modstand-flow" skema. Denne skematiske struktur spiller en helt afgørende rolle ikke blot i vores hverdagsopfattelse af fysiske fænomener, men også i fysikkens grundstrukturer i forskellige discipliner. Der er bl.a. derfor god grund til at tro at kropslige øvelser fulgt af nøje begrebslig analyse af studerendes erfaringer kan fungere som en god introduktion til fysikkens grundbegreber, og bygge bro over det fænomenologiske gab der er mellem den oplevede og den begrebslige forståelse. Vi giver eksempler på sådanne kropslige øvelser og beskriver hvordan øvelserne kan indgå i undervisningsforløb der sigter på at udvikle de studerendes forståelse og intuition i forhold til centrale fysiske grundbegreber.

**Introduction**

Physics teachers have long been using activities where students either directly feel or represent physical entities in instruction. For example, using students to illustrate the principles of refraction of light, how electrons move in a circuit (eg. Singh 2010), or how atoms behave in a gas, fluid, or solid (eg. McSharry and Jones 2000). Often, these activities are seen as illustrations that make the endeavour of learning physics more bearable or fun. Models are often meant to *illustrate* the underlying mechanisms of a physical phenomenon rather than having students actively link their experiences to a conceptual understanding of the laws of physics.

Such activities have also been described in physics education literature, and here they have many names: Participatory simulations (e.g. Collela 2000), analogical roleplaying (e.g. Aubusson and Fogwill 2006), drama (Ødegaard 2003), Energy Theatre™ (Sherr et. al 2013), and kinesthetic learning activities (e.g. Begel et. al 2004). In one case, students who had received instruction with kinesthetic models performed better than students that had not (Richards 2012). Common to all these activities is that students enact physical entities in order to understand scientific phenomena and concepts. In some cases, kinesthetic activities are seen as embodied forms of analogical models (Aubusson and Fogwill 2006). Authors emphasise that most students are able to use the activities to engage with each other in content related discussions and that students see this engagement as positive (e.g. McSharry and Jones 2000). This means that both from a cognitive and from a social aspect, kinesthetic activities and models can have a positive effect on student learning and classroom environments. However, researchers have not developed theoretical arguments based on cognitive or epistemological perspectives that justify and motivate using kinesthetic learning activities in physics teaching. Developing such an argument and illustrating some of the implications for teaching is the purpose of this article. We will focus the argument on the use kinesthetic learning activities in physics (primarily mechanics) teaching in secondary and tertiary education.

The argument is based on insights concerning embodied cognition and metaphor theory, conceptual change research and physics education research. The starting point is a closer consideration of the nature of the students' preconceptions and their connection to how physics learning can be viewed from the perspective of embodied cognition. After that we show how embodied cognition allow us to consider kinesthetic models as viable parts of learning activities that shape student understanding of mechanics. Specifically, we argue that the image schema *effort-resistance-flow* lies at the root of many areas of physics and of students' everyday observations. We then identify three types of kinesthetic models two in which effort-resistance-flow is crucial for linking student experiences with formal physics knowledge. For a particular kinesthetic model, we analyse student generated examples of such linking. Finally, we use Didactical Situations (Brousseau 1997; Ruthven et. al 2009) to structure the insights from the rest of the paper and present a design for teaching mechanics.

**Student conceptions of physical phenomena**

When students engage in an introductory university mechanics courses in physics, they are supposed to achieve a so-called Newtonian understanding of the mechanical world. A vast amount of literature has documented that traditional lecture based instruction is insufficient to achieve this goal (see e.g. Halloun and Hestenes 1985; McDermott 1991). Often students adopt surface approaches like rote memorisation and pattern recognition, rather than developing conceptual understanding of the subject. While surface approaches may enable students to pass the exam, only a minority will be able to use their physics knowledge outside a very narrow set of tasks. Students exposed to traditional lecture based mechanics instruction generally do not end up with a Newtonian understanding of the mechanical world surrounding them (Crouch and Mazur 2001).

Changing student conceptions from their "everyday" conceptions to the desired understanding of mechanical phenomena has been addressed extensively in conceptual change research (e.g. Hestenes 2006; Vosniadou 2010; Gentner and Colhoun 2010; McDermott 1991). A common theoretical assumption in this research tradition is that students are assumed to possess certain cognitive structures, schemas, which shape their actions and predictions. In the context of an educational setting, these schemas are triggered by a task perceived by the student and enacted in order to solve that task. What kind schemas of are triggered in a given situation and how they are enacted is believed to be influenced by a student's conceptual framework (Vosniadou 1994) or ecology (diSessa 2002). One of the central questions in conceptual change theory seems to be whether a student's conceptual understanding is stable across contexts, and thus theory-like, or whether different contexts is likely to trigger different schemas even if the formal theory around the task is the same. For example, will students give the same answer to a task involving the direction of acceleration of a point mass around a much larger mass, in the context of planetary motion as in the context of a child swinging a rock by a string?

It is a common finding in the conceptual change literature that conceptual change does not come about easily. A common teaching implication is that students' conceptual understanding need to be addressed and discussed explicitly in teaching, that teachers need to take students prior understanding seriously if teachers are to help students achieve Newtonian conceptions of mechanical phenomena. Even when taking preconceptions into account and addressing them in discussions, students do not necessarily adopt the formally correct views in their everyday lives. The problem of making students use Newtonian thinking outside a narrow context remains.

The idea of a conceptual ecology with regards to schemas (diSessa 2002) lends itself to a way of thinking about education where schemas and conceptual relationships survive based on how viable they are to a person or a collection of persons. Sumara and Davis (2006) describes this view as complexity thinking, in which the conceptual ecology of a person can be seen as nested in the setting of a classroom which in turn is nested in society. However, the point of teaching Newtonian mechanics must be that it be used outside the classroom - at least in other physics contexts. Thus, formally correct notions of Newtonian mechanics must me made viable outside the classroom, if they are to ever to be nested in other contexts.

While there is every reason to condone the efforts made in the literature on conceptual learning of physics with respect to taking seriously student common sense beliefs about mechanical phenomena (McDermott 1991), there is reason to question the robustness and consistency of students' misconceptions. As pointed out by diSessa (1993), the view that students have misconceptions that are quite stable and which must be confronted, is not only at odds with the basic idea of constructivism (that students should construct their knowledge on the basis of what they already know), but also not necessary to explain the experimental findings of the conceptual change literature. Another way of stating this point is to ask, what the epistemic status of these misconceptions is supposed to be. Are they coherent theories or are they rather ad-hoc hypotheses?

In an overview of conceptual change theories Özdemir and Clark (2007) describe two different trends in conceptual change research. The first, labeled *knowledge-as-theory* (e.g. Vosniadou 1994; Chi 1992; Posner et. al 1982), views students' conceptions are roughly described as theory-like structures, which provide the students with relatively unified, robust, and coherent views rooted in persistent epistemological beliefs. From the second perspective, *knowledge-as-elements* (e.g. diSessa 1983; Brown 1995; Clark 2006), students' mental structures are seen as less coherent with "mulitple quasi-independent" elements loosely connected and brought together in a more ad-hoc manner. The distinction is important because it has direct implications for instructional approaches. For instance, from the knowledge-as-theory perspective establishing "cognitive conflicts" may be aimed at in instruction, which may lead the student toward fundamental change of mental structures. From the knowledge-as-elements perspective, the question of cognitive development is one of refinement of existing mental structures rather than replacement of them. In our view changes in knowledge-as-elements are made to happen more incrementally than in knowledge-as-theory.

Özdemir and Clark argue that recent studies have provided empirical evidence for the knowledge-as-elements perspective. In addition, other studies from physics education research can be seen as indicative of the problems related to the knowledge-as-theory perspective. For instance, some studies show that the same students do not use the same misconceptions consistently. Thus, Hestenes (2006) reports that nearly all students are inconsistent in applying the same concept in different situations. The premise for solving conceptual problems correctly, is that the student is familiar with the general law and deductively applies it to the situation at hand and reaches a conclusion. However, if the student does not know which law to use or does not have knowledge of the general law to be used, the problem becomes one of establishing a relationship between the situation at hand, expected outcomes and an ad-hoc generated hypothesis about motion. Most students will likely be well aware that the general hypotheses about motion arrived at in this abductive fashion are highly fallible. Thus, the "misconceptions" are constructed in an ad-hoc manner as responses to the encountered conceptual problem. It therefore seems unlikely that students actually entertain e.g. 'Aristotelian' theories of motion and apply them to conceptual problems. Rather, they (may) arrive at 'Aristotelian' notions of motion as a result of (abductive) inferences when facing puzzling conceptual problems.

This way of looking at misconceptions is dynamic. Student conceptual systems vary between contexts, and one cannot be sure that the same student will make the same reasoning in problems that seem alike but have different contexts. Davis and Sumara (2006) argue that conceptual systems are not only complex but they consist of systems are themselves part of larger systems. Systems are nested within each other and their dynamics unfold on different timescales. In this view some conceptual structures would change over different timescales and due to different influences than other conceptual structures. The point is to find out what structures can be affected by different kinds of teaching. Our proposition is that influencing student conceptual systems through systematic use of their phenomenological and kinesthetic experiences in physics teaching could facilitate the emergence of conceptual structures that are (1) more stable than traditional teaching, and (2) aligned with formal physics.

### The role of embodiment when learning physics

But the problem remains – many students actually have severe difficulties at adopting "newtonian conceptions" of force, velocity, acceleration etc. How is it that some students arrive at the right Newtonian conclusions while others do not, if it is not so that students seriously entertain faulty theories about motion? The answer to this question is, of course, extremely complex. Andrea diSessa argues, as discussed briefly above, that understanding this question involves moving towards a systems perspective of knowledge, leaving behind the simple "misconceptions schemas" and replacing this notion with a multitude of mental substructures and diverse knowledge forms invoked in different situations (diSessa 1993, pp. 148-149). Notable among these are the so-called p-prims or phenomenological primitives, which are used in the construction of mental models. The term 'primitive' is used to designate that the p-prims are used as if they needed no explanation – they provide with intelligibility of phenomena (or provide structure to phenomena), if not explanation (diSessa 1993, p. 112). These primitives are related to different types of phenomena familiar from everyday life, for instance phenomena pertaining to force and agency (e.g. "dying away") or constraint phenomena (e.g. "blocking).

P-prims bear strong resemblances to a type of embodied schematic structures discussed in cognitive linguistics: The image schemas (diSessa 1993, p. 122). These embodied schemas, suggested by Lakoff and Johnson (1983), play a crucial role in metaphorical reasoning. While diSessa's p-prims are typically (but not necessarily) anchored in and derived from bodily experience (diSessa 1993, p. 122-23), image schemas are necessarily so anchored.

**Metaphor, image schemas, p-prims**
Over the past 30 years the importance of metaphor and analogy in learning processes have been stressed by many authors (Gentner and Colhoun 2010; Lakoff and Johnson 1980; Johnson 1987; diSessa 1983; Lakoff 1987). Basically, the idea of learning by analogy is that the learner has a good understanding of the conceptual topology of a base domain and maps this conceptual structure unto a less well understood target domain (Lakoff 1993), thereby providing conceptual structure to the target domain.

According to Lakoff and Johnson (1983) metaphors get their conceptual structure or topology through image schemas (see also Lakoff 1987; Johnson, 1987). Image schemas are pre-linguistic structures acquired through our basic bodily experiences. Lakoff and Johnson (1983) discuss several such rudimentary schemas, for instance such schemas as "part-whole", "source- path-target", "up-down", "center-perifery" and others. As can be gathered, these image schemas are extremely general, and serve as phenomenological building blocks of our cognition, and the foundation for analogy and metaphor: Two conceptual domains may be seen as similar with reference to a common image schematic structure. Thus, for instance, the conceptual metaphor "Life is a journey" lends is basic conceptual topology from the *source-path-target* image schema which has a bodily basis. This basic schema provides structure to the concept of "journey", and is projected unto "life".

One of the most important p-prims discussed by diSessa is the so-called "Ohm's p-prim", a p-prim that is absolutely crucial in making sense of mechanical phenomena. This p-prim concerns "an agent that is the locus of an impetus that acts against a resistance to produce some sort of result" (diSessa 1993, p. 126). Thus, this p-prim surely has a bodily basis – the act of pushing material objects, experiencing resistance and causing a resulting movement (diSessa 1993, p. 126; see also Hestenes 2010, p. 28). As Hestenes (2010) puts it: "No doubt it [Ohm's p-prim] originates in personal experience of pushing material objects, and it is projected metaphorically to other situations".

From the phenomenological perspective (i.e. the way we experience phenomena around us) one of our most fundamental experiences is the bodily experience of exerting effort, experiencing resistance and giving rise to flow. Johnson's description of force-based image schemas can be seen as aspects of effort-resistance-flow; they may be understood as being special cases of effort-resistance flow. For example, "blocking" (Johnson 1987) may be seen as effort against a resistance with no flow. Thus, this rudimentary conceptual structure plays a crucial role in our everyday movements and interactions with the world around us. It is also easy to see how this structure may lead us to false conclusions about the nature of motion. For instance, a common "misconception" referred to in the "misconception" literature is the idea that heavier objects fall faster than light objects. The strange thing about this idea is that it is so easy to demonstrate that it is not the case. So how do many students arrive at this misguided conception so easily refuted by experience? A likely explanation is that many students are making a metaphorical mapping from bodily experience using the effort-resistance-flow schema. Students know from their own bodily experiences that lifting a heavy object requires more effort than lifting a lighter one, and reason (wrongly) that, conversely, the heavier object will fall faster when falling towards the earth. The excerpts from Halloun and Hestenes (1985) illustrate vividly that many students simply do not have the conceptual apparatus needed to distinguish precisely what is going on in such fundamental situations. They know the feeling of difference between heavy and light objects, but their conceptual understanding of concepts of acceleration, force and mass are not sufficient to make inferences that are consistent with Newtonian thinking.

Halloun and Hestenes refer to Buridan's description of the concept of "impetus" and argue that this theory is very much like the "vague intuitions common among students": "A mover, while moving a body, impresses on it a certain impetus, a certain power capable of moving this body in the direction in which the mover set it going [..]". This, in a nutshell, is the *effort-resistance-flow* schema, but - as is crucial to our argument - *effort-resistance-flow* can be cued (or activated) in other contexts as well. For example, in particular contexts the expressed idea that motion must have a cause (Hestenes et. al 1992) can be associated with this schema, resulting in "Aristotelian thinking". We think that students' "misconceptions" stem from imprecise conceptual understandings of physics concepts (see Figure 1). But students are completely familiar with the phenomenological experience of these concepts in action. In line with diSessa's (1993) thinking on p-prims, we believe that image schemas or the way in which they are cued, can be aligned with Newtonian mechanics. The task for instruction is then to align phenomenological experiences with formal physics concepts and relations.

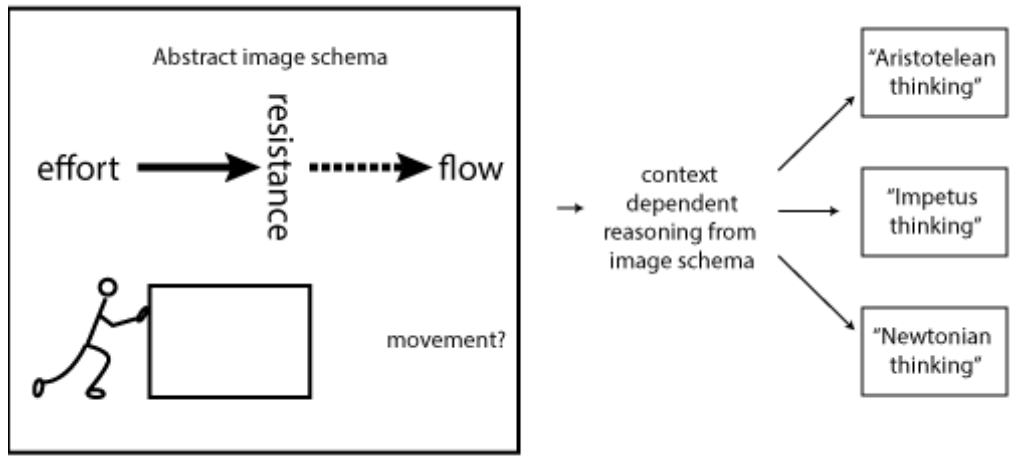

*Figure 1. An illustration of the image schema effort-resistance-flow. We claim that since effort-resistance-flow is ubiquitous, it may be the source of many different types of "misconceptions", depending on the context in which it is used.*

**Effort-resistance-flow as a central topology in the structure of physics**

The effort-resistance-flow schema is not at odds with physics but lies rather at the heart of physics. In many physical disciplines the central physical variables either drive motion or describe it (Hogan and Breedveld 2008). Thus contemporary energy bond graph theory used in engineering considers effort-variables (e.g. force, voltage, hydrodynamic pressure) and flow-variables (e.g. velocity, current, volumetric flow rate). The relation of these variables signifies an "energy bond", i.e. a power transmission. Energy bond graph relies on an analogy between physics domains that first developed by Maxwell and has been developed in the methodology of energy bond graphs (Paynter 1960). Table 1 lists examples of effort and flow-variables along with pictures of prototypical systems.

Table 1. Examples of variables that are used in systems modeling in engineering and proposed relation to the effort-resistance-flow image schema. In each of these systems, we can identify effort and flow variables and also resistances.

| Domain | Effort (symbol [SI unit]) | Flow (symbol [SI unit]) | Resistance | Prototypical system |
|---|---|---|---|---|
| Linear mechanics | Force (F [N]) | Velocity (v $[\frac{m}{s}]$) | Mechanical resistance | |
| Electric circuits | Voltage (U [V]) | Current (I [A]) | Electrical resistance | |

| Hydrodynamics | Pressure (P [Pa]) | Volumetric flow rate (Q [$\frac{m^3}{s}$]) | Pipe resistance (Lautrup, 2011) | 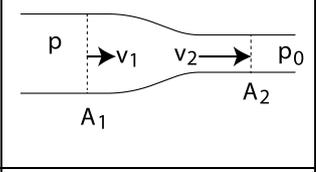 |
|---|---|---|---|---|
| Embodied | Kinesthetic sensation of applying effort | Kinesthetic sensation of flow | Felt resistance | 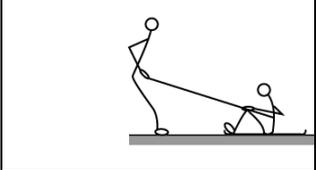 |

While effort-resistance-flow is a central image schematic foundation for physics, many other image schemas may influence conceptions of physics. For example, containers (eg. using Gauss boxes to calculate electric fields), end-of-path (ray diagrams in optics), scale (eg. units), cycles (planetary motion), and balance (e.g. statistical mechanics) may all be relevant. However, effort-resistance-flow is likely central to the sense of mechanism (diSessa 1993) experienced by people successfully performing and those learning physics, because it is tightly connected to our being able to interact with entities in the world.

This fundamental insight should be utilized in teaching by devising experiments and teaching that explores the analogy between the students' basic bodily experiences and physical concepts in various disciplines. Our being in the world shapes our language and concepts via image schemas in what appears to be an evolutionary way. That is, with somewhat random variations and selection based on best fits to the current situation. Then teaching should be able to use students' being in the world to target their language about and conceptions of physics. Such teaching would be dependent on designed situations where explanations that are consistent with physics language and conceptions provide the best fit to the situation.

**Different types of kinesthetic models**

Energy bond graphs exploit underlying common effort-flow structures of different areas of physics using a diagrammatic form of representation (see Hogan and Breedveld 2008). Using that line of thinking, we propose that kinesthetic learning activities can exploit relevant image schemas in connecting human experience with, for example, diagrammatic or verbal forms of representation.

In one type of model, students use *effort-resistance-flow* as a basis for relating their experiences in a model to a formal physics description these experiences. In this sense, the source domain is really the same as the subject domain: Students enact being physical objects (see Figure 2) and the target domain is Newtonian physics. However, since the students really are mechanical objects, there is no material difference between the two domains: Newtonian mechanics describe motion in the human domain.

Another type of model students use *effort-resistance-flow* to model interactions of some model or theory. Students enacting being electrons in a circuit (Singh 2010) may use *effort-resistance-flow* to connect the push from other students to the effect of voltage in the circuit. The working hypothesis of such models is that it is possible to make a mechanical kinesthetic model with the same types of relations between variables as there would be in an electrical, thermodynamic, hydrodynamic, or quantum mechanical system

In a third type of model, students may use other image schemas (such as *containment*) to model other aspects of a model or theory. One could argue students in Energy Theatre™ (Sherr et. al 2013) make use of, among other things, *containment* and *conduite* metaphors to embody their understanding and discussions of energy transfer processes. In the last two types of models, the source domain (mechanics/phenomenological experience) is different from the target domains (electrical circuits and energy transfer).

In neither of the types of models are the students utilised as sensors (Sherr et. al 2013) that merely detect some physical quantity, nor are they to perform movements to stimulate the brain to learn better (Hannaford 1995). Rather they should use their experiences with enacting the models to discerne physics concepts. For example, if a student pulls another student with a rope, they will both feel pressure from the rope and tension in their bodies, even if they do not identify the pull as a force in a Newtonian way. The point is to relate these bodily experiences to formal physics, including generalizing some experience and downplaying experiences that are not related to the abstractions that teaching intends students to make.

**A model relevant for teaching mechanics**

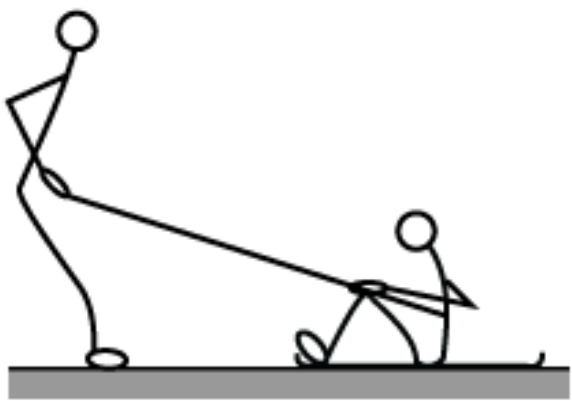

*Figure 2: Drawing of the kinesthetic experiential model discussed below. One student is sitting on a slab of plastic while being pulled by another student. In practice, some students hold on to the rope directly, while others attach it to the slab.*

In this section, we provide an example of a kinesthetic model (Bruun 2009; Bruun and Johannsen 2014) designed to work Newtonian mechanics, in particular, linear motion (see Figure 2). Students may use kinesthetic experiences as a basis of written accounts of their experiences of effort-resistance-flow, and below are examples from upper secondary students with little prior knowledge of Newtonian mechanics. Students had participated in a teaching unit in which they were asked to perform kinesthetic learning activities[1] while at the same time discussing and writing down (1) descriptions of their kinesthetic experiences, (2) physics concepts they believed were relevant to their experience, and (3) explanations that related experiences with concepts. Examples of these writings are presented in Table 2 and 3 below. Bracketed text, [text], represents the authors' extrapolation of student accounts given

---

[1] We report on enactments of this teaching unit elsewhere (Bruun and Johannsen, 2014).

knowledge of the specific teaching context.

*Table 2. An example from a student describing the kinesthetic experience of pulling/being pulled while sitting on a slab while holding on the rope. The top row is in Danish, while the bottom row is in English.*

| Kinesthetic experience | Physics concept | Explanation |
| --- | --- | --- |
| Jo flere kræfter desto hurtigere gik det. [Det blev] lettere [at trække]. | Kraften går gennem snoren. G-kraften. | Man skal starte med at løbe hurtigt for så bliver det også hurtigere lettere [at trække]. |
| The more effort [forces?] the faster it went. [It became] easier [to pull] | The force goes through the string. G-force. | You have to start out running really fast because then it quickly gets easier [to pull] |

The example in Table 2 clearly reflects to topology of effort-resistance-flow. More effort begot more movement. The physical concept of (G-)forces acting through the string is imprecise, and the explanation also follows the effort-resistance-flow topology (the felt resistance is reduced if the flow increases quickly). It is tempting to categorise the example as an impetus misconception. However, the explanation is not wrong. A formal analysis of the situation shows that the force needed on the box/sitting person to sustain a constant velocity is independent of the velocity. However, increasing the velocity would require a larger force. Thus, while accelerating the student will feel more resistance than when a constant velocity has been reached. If the constant velocity is reached quickly the reduced resistance will also be felt quickly. Clearly, the student in the example does not connect his own experiences with formal physics in the manner just explained, and we would not expect most students to do this. This is the teacher's job: Focussing on students explanations as a base domain for connecting to the domain of physics. The teacher can point to things in the explanation that should be generalised, help students map their experiences to the language of formal physics, and make suggestions for variations that help

*Table 3. A student description of the kinesthetic experience of pulling a slab. The top row is in Danish, while the bottom row is in English. The Danish word* kræfter *kan be used both with a physics meaning and an everyday meaning close to* effort.

| Kinesthetic experience | Physics concept | Explanation |
| --- | --- | --- |
| Begge skal give kræfter for [at] man [kan] holde balancen. "Hiveren" skal dog bruge flere kræfter. | Modstandskraft | Når jeg trækker er det hårdest at få kælken i gang. |
| Both need to provide | Resistance force | When I pull, getting the |
| [effort] so [that] you can keep the balance. However, "the puller" needs to use more [effort]. | | slab going is the hardest. |

Table 3 shows a case where effort is connected to balance. But the puller needs to use more effort. This example could be interpreted as a misconception, where the puller provides the most force since the puller is active while the pulled person is regarded as passive (see Hestenes et. al 1992). However, the puller *does* need to provide more force than the sitting student to keep the system moving. He provides tension to the rope that is large enough to accelerate/sustain motion of the sitting student. The sitting student also provides this force. In addition the pulling student exerts a force on the floor to create a reaction force that will accelerate and then sustain the motion of the system. A teacher could help students generalise the descriptions to Newton's third law, by focussing on the interaction between the puller and the pulled, and downplaying the puller's interaction with the floor.

As is evident, students are not likely to miraculously derive important physics concepts simply by performing kinesthetic learning activities. Guiding questions, validating reasoning, and timely explanations are essential elements of the teacher's toolbox as in any kind of teaching. However, due to the novelty of kinesthetic models in teaching, we use the next section to illustrate how one can plan this kind of teaching. We use the Theory of Didactical Situations (Brousseau 1997) to frame teaching that uses the model from this section (Figure 2).

**Framing physics teaching by using kinesthetic learning activities**

The Theory of Didactical Situations (TDS; Brousseau 1997) is a design and analytical framework developed for mathematics. However, it has proven usable for a wide variety of subjects, including physics (Ruthven et. al 2009; Tiberghien et. al 2009). Ruthven et. al (2009) describe in detail, how TDS as an intermediate framework allows designers to incorporate abstract elements from grand theories of learning in teaching designs that can be realised in a classroom. TDS is in its original form heavily influenced by Bachelard's (1938; 2002) notion of epistemological obstacles and Piaget's (Gruber and Vonèche 1995) notions of disequilibration and accommodation. In our case, the grand theory corresponds to conceptual metaphor, and we incorporate the abstract notion of image schemas, in particular *effort-resistance-flow*, through kinesthetic enactments of the pulling model described in the previous section.

We proceed to use TDS to structure a lesson around that model. TDS revolves around the concept of adidactical situations where students "are engaged directly with solving a novel type of problem, refining their concepts and strategies in the light of feedback from a (material and social) milieu" (Ruthven et. al, 2009). There are three crucial elements to an adidactical situation: A *task* that will have the students autonomously engaged with the knowledge to be learned. In this example, the central task is *to explain kinesthetic experiences in ways that are consistent with formal physics* (see Supporting Material for a full lesson plan including worksheets). The second component is comprised by the *conditions* under which the task will be solved. In this lesson, materials such as rope and slabs, worksheets, student groups of three, and an open setting with multiple surfaces comprise these conditions. Finally, an adidactical situation will need to specify an *expected progression* towards the knowledge to be learned. In this lesson it is for students to enact and discuss the model several times in order to map their kinesthetic experience of effort-resistance-flow to a concept of force that is consistent with Newtonian mechanics.

To effectuate movement when pulling another student, the puller needs to use some parts of their bodies to maintain tension in the rope and other parts to produce movement. For students it may be part of the same experience, but separating the two sensations could be a useful step to analyse the situation from a physics point of view. Change in movement (flow) would be the result of (1) a force from the floor and (2) tension in the rope acting on the student. The assumption made in the planning of the lesson is that separating the different sensations should lead to students being able to (1) map effort in the legs to a force from the interaction with the floor, while (2) effort in the arms and torso is related to a pull on the other student[2]. Successful mapping of a single experience of pulling to two separate sensations can be viewed a process of accommodation. Taking this view, the preceding disequilibration should be facilitated by a task given on the worksheet which explicitly asks students to consider which parts of their bodies are used in the movement. To validate student mappings, concepts, and explanations, the *didactical environment* - that is the place in which the adidactical situation takes place in combination with the teacher and students that are engaging in the situation at that time - must enable students to evaluate their own thinking.

The designer makes deliberate choices about the means by which the students can do this self evaluation. The materials (e.g. rope, handles, and slabs), the worksheets, group sizes, and even the setting (including how other students and the teacher act) comprises *didactical variables* that can be adjusted, added or removed. For example, in this lesson groups of three emphasise the focus on three particular roles: one being pulled, one pulling and one observing. This choice was made to help students compare and contrast different experiences when making explanations. The intention is that student dialogue should help lead to disequilibration and subsequently accommodation of, for example, the notion of splitting up the sensation of pulling. The mode of the teacher is to ask questions that facilitate dialogue. Another example of didactical variables are the materials, which have been chosen to allow different ways of enacting the model. Thus students can investigate different ways of pulling/being pulled and different surfaces. These are conscious choices designed to support students' development of explanations that are consistent with physics.

From the designers point of view the adidactical situation should foster student knowledge - explanations of their kinesthetic experiences - that is recognised by the students as the most viable solutions to the tasks. Students successful in solving the task should be on the way to bridging, what we term here, the *phenomenological gap* between everyday experiences and the abstractions of formal physics. Figure 3 shows in schematic form how a designer/teacher can plan for such situations. The actual teaching situation is necessarily concrete and contextualised. Following

---

[2] In a force diagram, the force on the pulling student originates from the object being pulled. The law of action/reaction could be described as another situation with disequilibrium and accommodation.

Brousseau (1997), the knowledge to be learned was originally created in a specific context, as in some of Newton's experiments, but then decontextualised, as in F=ma. A designer of teaching must re-contextualise it, in this example by creating a kinesthetic model and planning for kinesthetic activities. This process of planning is usually referred to as a transposition; the abstract model from physics is changed and positioned relative to the teaching context. Following the curved arrows going to the right in Figure 3, the designer adds context and thus also makes the situation more complex. The end of the path is the actual teaching situation. The job of teaching and learning is then to have students re-decontextualize the knowledge to an abstract form that is consistent with, in this case, Newtonian mechanics. This is illustrated in Figure 3 with curved arrows going to the left. We propose that this is a way of teaching with kinesthetic learning activities can help students integrate relevant image schemas, which have impact in their everyday lives, with the abstract representational forms of formal physics.

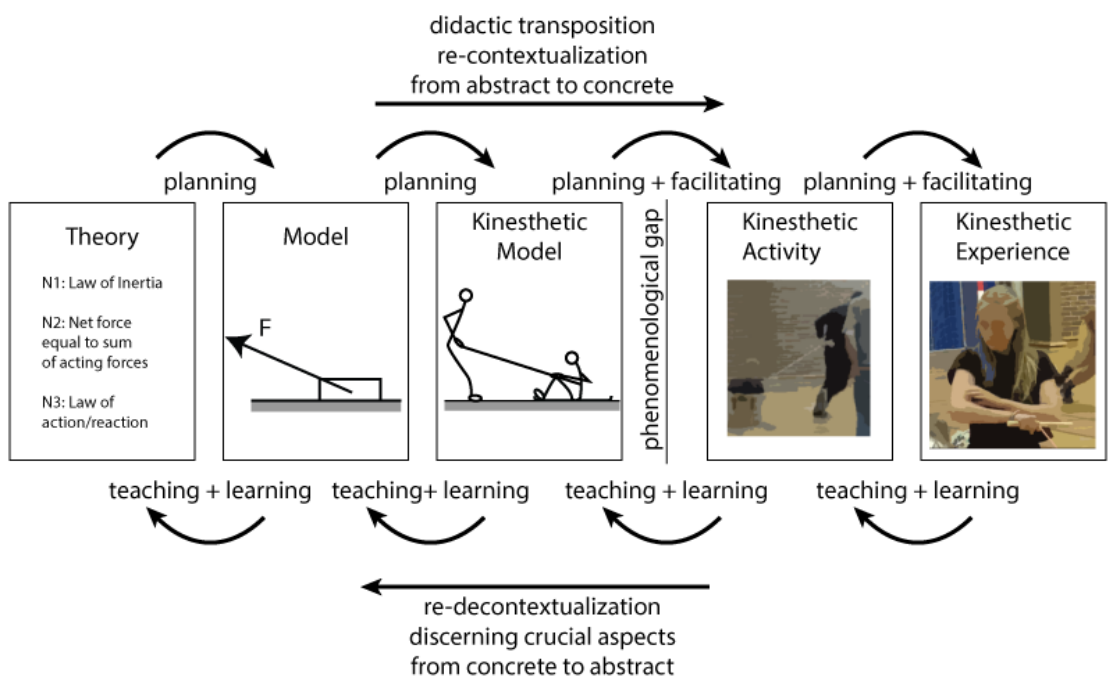

*Figure 3: Levels at play in planning and teaching with kinesthetic models and exercises. Central to the lesson is bridging the phenomenological gap between the kinesthetic experience and more abstract models.*

As is pointed out by Ruthven et. al (2009), it was found necessary to embed the adidactical situation in a didactical structure. First of all, students need to be empowered at the beginning. That is, the teacher distributes some of the responsibility and freedom of action to students by handing over the task and divulging the didactical environment. In this lesson, the teacher shows the equipment (ropes and slabs), and shows how to enact the model safely without prescribing how students should enact the model. In TDS, this phase is called the *devolution* (Winsløw, 2006).

Students then proceed to enact the model and engage with the task in *action phases* where they enact the kinesthetic model. They formulate their explanations verbally, in writing, and through drawings in *formulation phases*. Students validate their thinking via the feedback mechanisms in the didactical environment using other students, the teacher, or more enactments of the model in *validation phases*. These phases are all part of the adidactical situation. An initial action phase precedes formulation and validation phases in this particular lesson, but after the initial action phase, the three phases are intertwined forming small "eddies" of action-formulation-validation (e.g. enacting-drawing-discussing).

In a final and crucial *institutionalization* the teacher (or the students) contextualises the intended knowledge. The idea is to connect the knowledge that has been developed during the lesson to the official or institutionalised knowledge. In this lesson the final phase of institutionalisation is comprised of a classroom discussion, where the teacher highlights the generalizable aspects of student experiences in the didactical environment, while downplaying aspects that are specific to the environment. We summarize this description in Table 4.

Table 4: Five phases of TDS describing a lesson that integrates the pulling model. The first column gives a short explanation for what occurs in each phase. The second column show student roles, and the final column exemplifies the two former for the lesson described here.

| Phase: short explanation | Students role | Example from lesson with pulling model |
|---|---|---|
| *Devolution:* Hand out material unfold the didactical environment and task. | Taking ownership of activity. Need to figure out how they actually enact the models. Becoming aware of what they need to pay attention to. | Students might try attaching the rope around their waist, holding on to the rope with their hands, or other ways of enabling the pulling motion. |
| *Action:* Students enact models to focus on what kinesthetic sensations they experience. | Enact the models and change roles (eg. being pulled, pulling, observing). Testing hypotheses. | Some students run fast, others walk slowly. One hypothesis is that this has effect on the effort experienced to maintain the movement. |
| *Formulation:* Put words and images to kinesthetic experiences and come up with (physics) explanations | Discuss their ideas using their experiences with the models from different perspectives (eg. pulled, pulling, observing). Formulate these discussions on paper. | See examples from Tables 2 and 3. |
| *Validation:* Explanations are validated by enacting the models and by student-student and teacher-student interactions. | Enact kinesthetic models to validate their explanations and discussions. Use other strategies for validating observations (calculations, textbook knowledge, etc.). | If all three students try running fast and slow, they can discuss their kinesthetic experiences. They might ask the teacher as well, although this could easily become a premature |

| | | institutionalization. |
|---|---|---|
| *Institutionalization:* Teacher uses classroom dialogue to connect kinesthetic experiences to physics. | Speak of kinesthetic experiences increasingly in terms of forces and acceleration. | Teacher highlights that pull on hands should feel the same regardless of speed, while downplaying differences in speed between students. |

The goal of with this lesson is for students to inductively, abductively[3], and by ways of analogy to use the kinesthetic model to reach a higher level of abstraction than the kinesthetic activity itself. Students should be able to use abstract physics concepts, like force and acceleration - and at some point also representational forms like force diagrams and equations - to solve more general problems and give explanations consistent with formal physics. In order to do that, teaching must be designed to bridge the phenomenological gap between physics learners' kinesthetic experiences and abstract physics knowledge.

By discussing, drawing, and writing down their own ideas, students use multiple forms of representations to formulate their knowledge as they are constructing it. This knowledge must stand the test of being the most viable solution to the task given the didactical environment. Promoting continuously that students enact the models as part of their discussions, in between writing down and drawing, and even as part of validating their own explanations, the kinesthetic activities are likely to afford the creation of close connections between formal physics and phenomenological experiences. And it is by promoting enactment, discussion, validation, and subsequent institutionalization of knowledge that we expect image schemas to be modified to resonate more with physics, perhaps in the sense of cuing discussed by diSessa (1993).

**Concluding remarks**

We have developed an argument for using kinesthetic activities in instruction that focuses on changing physics conceptions of students. We have argued that image schemas provide a fruitful entrance to facilitate this change, and that in particular the effort-resistance-flow schema is central to most school physics. We have been informed by literature to identify different types of kinesthetic exercises and we have specified how one may design instruction that affords conceptual change.

We hold that it makes little sense to label student understandings as right or wrong. In our view, (image) schemas as enacted by students have served them well (mostly) in their lives so far, a point made also by Linder (1993). To initiate change in student conceptual ecologies (diSessa, 2002) there must be a designed situation that selects for understandings consistent with formal physics. Using evolution as a metaphor (Sumatra and Davis 2006), the task for designers, teachers and students is to breed conceptual ecologies where Newtonian mechanics are fit enough to survive and even thrive. Such ecologies should also be context aware, so they can be used appropriately by students (Linder, 1993).

---

[3] Ascribing a kinesthetically experienced effect, like a tingling in the stomach to an increased acceleration might have to be an abductive process. Other things could have caused the tingling.

The specific image schema that we have termed effort-resistance-flow is likely an ideal starting point for learning physics, as most school physics can be encompassed by the schematic structure as demonstrated by energy bond graph theory. A central aim of instruction from this perspective is that students become aware of how their intuitive experience of effort-resistance-flow situations may be conceptualized and used to work with and explain physics phenomena. Kinesthetic activities are based in the same domain as everyday experiences and these experiences are believed influence image schemas. If image schemas are indeed central to human development of conceptual understandings, the idea that instruction might target them through kinesthetic activities should be explored further.